\documentclass[prl,twocolumn,showpacs]{revtex4}
\usepackage{graphicx}
\begin{document}
\title{Multistable transport regimes and conformational changes
in molecular quantum dots}
\author{Alexander O. Gogolin and Andrei Komnik}
\affiliation{Department of Mathematics, Imperial College, 180 Queen's Gate,
London SW7 2BZ, United Kingdom}
\date{\today}

\begin{abstract}
We analyse non-equilibrium transport properties of a single-state
molecular quantum dot coupled to a local phonon and contacted by
two electrodes. We derive the effective non-equilibrium (Keldysh)
action for the phonon mode and study the structure of the saddle
points, which turn out to be symmetric with respect to time
inversion. Above a critical electron-phonon coupling $\lambda_c$
the effective potential for the phonon mode develops two minima in
the equilibrium and three minima in the case of a finite bias
voltage. For strongly interacting Luttinger liquid leads
$\lambda_c=0$. Some implications for transport experiments on
molecular quantum dots are discussed.
\end{abstract}

\maketitle
Recently the progress in engineering and production of
nanoscopic electronic devices reached a new summit as
experimentalists reported successful contacting of single
molecules thereby opening the way to build molecular quantum dots
\cite{c60,weber}.
Since bistable FET-based
elements play the most important role in modern microelectronics a
fabrication of such devices on a nanometre scale would be a
natural way of further progress.
The molecules involved in the electron tunnelling process
naturally possess elastic degrees of freedom which are bound to
respond to the applied bias voltage in some way; this has been
dubbed -- `conformational change' (see e.g. \cite{Kr}). There have
been numerous studies of similar systems \cite{Meir,Schoen} though
none of them has, to our knowledge, concentrated on the behaviour
of the internal degrees of freedom in the relevant
non-perturbative, non-equilibrium regime.

To understand the basic physics of the conformational change, in
this Letter we ignore all the mindboggling complexity of real
molecular dots \cite{kornilovitch} and employ the simplest
possible Holstein--type model with a single phonon mode
\cite{holstein}. We study how this phonon responds to the
non-equilibrium electron environment. If one applied a `maximum
entropy' principle (such as would be invoked to explain the
time-reversal breaking in driven \emph{macroscopic} systems), one
might expect that the molecule would deform is such a way as to
facilitate the current flow. By developing the full
non-equilibrium theory we shall show that such scenario, while
admissible in principle,  is not correct for the \emph{quantum}
system in question. Instead we find that in this case a
generalised non-equilibrium free energy is being minimised and
multistable transport regimes emerge.

\begin{figure}
\includegraphics[scale=0.22]{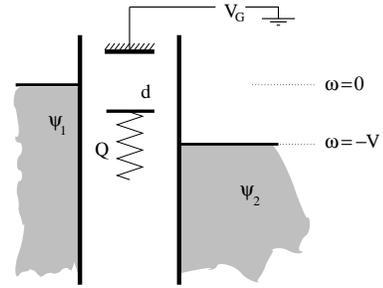}
\caption[]{\label{dotsetup} Schematic representation of the system
under investigation. A single quantum dot level $d$ coupled to a
local boson mode $Q$ can be populated via tunnelling from two
fermionic reservoirs $\psi_1$ and $\psi_2$. The latter are kept at a
fixed chemical potential difference $V$. The gate voltage $V_G$
enables controlling of the dot level population.}
\vspace*{-0.5cm}
\end{figure}

Let us begin with the Hamiltonian of the system,
\begin{eqnarray} \label{H0}
 H = H_0[\psi_1] + H_0[\psi_2] + H_Q + (\Delta + \lambda Q) d^\dag d + H_t \, .
\end{eqnarray}
The first two terms on the rhs describe non-interacting
Fermi seas in the left and right electrodes;
$\psi_1$, $\mu_1=0$ and $\psi_2$, $\mu_2=-V$ are the corresponding
electron annihilation operators and chemical potentials, $V$ is
the bias voltage, see Fig.~\ref{dotsetup}. The third term
describes the phonon mode , $H_Q = (M \dot{Q}^2 + K Q^2)/2$, where
$K$ is the elastic constant and $M$ is the oscillator's mass. The
quantum dot is described by the fourth term and contains only one
electronic state, which annihilation and creation operators are
denoted by $d$ and $d^\dag$, respectively. It can be occupied only
by one electron (the spin effects are unimportant as long as the
Hubbard coupling is weak so that the system does not enter the
Kondo regime, which we assume throughout), which is coupled with a
coupling constant $\lambda$ to the local phonon mode $Q$ and has
the offset energy $\Delta$. The latter can be controlled by
changing the gate voltage $V_G$, see Fig.~\ref{dotsetup}. The last
term is responsible for the transport through the dot and reads
$H_t = \gamma[ d^\dag(\psi_1 + \psi_2) + {\rm H.c.}]$, where
$\gamma$ is the energy independent tunnelling amplitude (we assume
that the leads are symmetric to simplify formulas).

\begin{figure}
\includegraphics[scale=0.25]{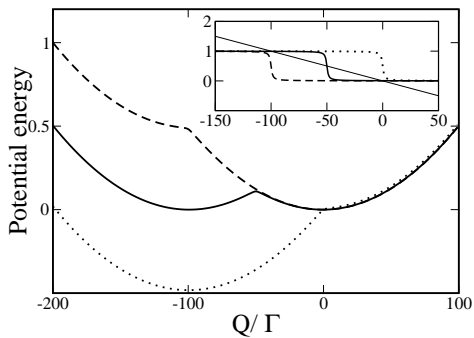}
\caption[]{\label{potplot} Effective potential energy of the
phonon as a function of $Q/\Gamma$ for three different gate
voltages corresponding to $\Delta/\Gamma = 0$, $50$, $100$
(dotted, solid and dashed lines, respectively). In the inset: the
graphical solution of Eq.~(\ref{findQe}). The straight line
represents the lhs with $K \Gamma/\lambda^2=1$. The curves show
the behaviour of the $n_d$ (rhs of Eq.~(\ref{findQe})) for the
same $\Delta/\Gamma$ as in the main graph.}
\vspace*{-0.7cm}
\end{figure}

\begin{figure}
\includegraphics[scale=0.25]{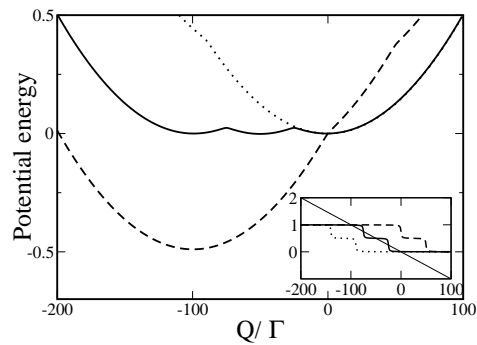}
\caption[]{\label{potplotV} Plot similar to Fig.~\ref{potplot} but
for the non-equilibrium case with bias voltage $V/\Gamma=50$. The
dashed, solid and dotted lines correspond to $\Delta/\Gamma = 0$,
$75$, $140$, respectively. The inset shows the graphical solution
of Eq.~(\ref{findQe}) for the non-equilibrium situation. }
\end{figure}

\begin{figure}
\vspace*{0.5cm}
\includegraphics[scale=0.24]{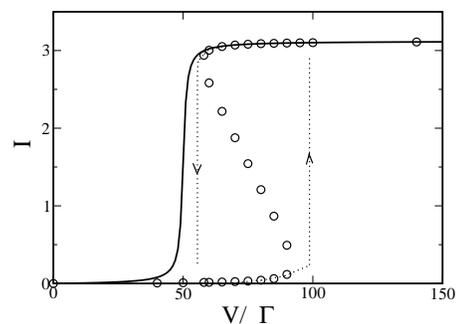}
\caption[]{\label{IV} Current-voltage diagram of a dot with zero
offset $\Delta=0$ as a function of the applied bias voltage $V$
(circles). The current $I$ is calculated via Eq.~(\ref{tokI})
after solving the Eq.~(\ref{findQe}) and is normalised to $e^2
\gamma^2/2 \pi$. The dotted lines represent the actual behaviour
of the current whereas circles in between them are the unstable
solutions of the Eq.~(\ref{findQe}). For comparison: solid line is
the $I-V$ characteristics of a free dot ($\lambda=0$) with a
finite offset $\Delta/\Gamma = 50$, $K \Gamma/\lambda^2=1$. }
\vspace*{-0.6cm}
\end{figure}

In most cases the molecular vibrational modes are slower
then the electronic degrees of freedom so that it is possible to work in
the Born-Oppenheimer (or static) approximation.
Within this approximation the
instant value of the current is given by:
\begin{eqnarray}     \label{tokI}
 I = \frac{e^2 \gamma^2}{2 \pi} \left[\tan^{-1}\left(\frac{\Delta +
 \lambda Q}{\Gamma}\right)-
 \tan^{-1}\left(\frac{\Delta +\lambda Q -V}{\Gamma}\right) \right]\,,
\end{eqnarray}
where $\Gamma=2\pi\gamma^2\rho_0$ is the resonance width
($\rho_0$ being the local electron density of states).
Outside the static approximation,
the phonon coordinate $Q$ still possesses its own dynamics.
Our goal is to derive the effective action for $Q$.
The obvious way to proceed is to eliminate the leads and dot
variables in the functional integral representation.
Since we are interested in the non-equilibrium situation,
we work in the Keldysh representation, where all time integrations are
performed along a closed path $C$ which consists of the
time-ordered branch $C_-$ and the ani-time ordered branch $C_+$
\cite{keldysh}.
Generalising the procedure of Ref.~\cite{yuanderson}, we define the
functional,
\begin{eqnarray} \label{Zdefine}
 Z[Q] = \langle T_C e^{-i \lambda \int_C dt Q(t) d^\dag(t)
 d(t) } \rangle \, ,
\end{eqnarray}
where $T_C$ stands for the contour ordering operation and the
average is taken over the exact eigenstate (the steady state) of
Hamiltonian (\ref{H0}) with $Q=0$. We set $\Delta=0$ (it will be
restored later) and divide the field $Q$ into two
Keldysh components scaled by $\lambda$: $Q_\mp = \lambda Q(t~\mbox{on}~C_\mp)$.
The effective action is then $S[Q_\mp]=S_0[Q_\mp]+i\ln Z[Q_\mp]$,
$S_0[Q_\mp]$ will be specified shortly.
One way to calculate the effective action is to take
the functional derivative of $Z[Q_\mp]$ respect to, e.g., $Q_-$:
\begin{eqnarray}   \label{Zvariation}
 \frac{\delta Z}{\delta Q_-} = -i
 \langle T_C d(t_-+0^+) d^\dag(t_-)e^{-i \lambda \int_C dt Q(t) d^\dag(t)
 d(t) } \rangle \, ,
\end{eqnarray}
$t_-$ lies on $C_-$.
If both sides of this equation are
divided by $Z$ in order to cancel disconnected
diagrams introduced by $Q_-$ being different from $Q_+$, then the
rhs becomes the time-ordered Green's
function of the dot level calculated in the presence of $Q_\mp(t)$,
\begin{eqnarray}   \label{Ddef}
 D^{--}(t,t') = - i \langle T d(t) d^\dag(t') \rangle \, .
\end{eqnarray}
We now implement the non-equilibrium generalisation of the
standard Born-Oppenheimer approximation by calculating the above
function (i.e. the electronic response to the phonon) for static
time-independent (but different) $Q_\mp$, restoring the time
dependence at the end of calculation. By means of fermionic
functional integration or via the method of Ref.~\cite{caroli} we
therefore obtain:
\begin{eqnarray}     \label{exactD}
 D^{--}(\omega) = \left\{  \begin{array}{rr}
                  \frac{\omega-Q_--i \gamma^2 \mbox{\small sign}
                  \omega}{(\omega-Q_-)^2+\Gamma^2} \, ; & \omega>0 \,
                  \mbox{and} \, \omega<-V
                  \\ \\
                  \frac{\omega-Q_+}{(\omega-Q_-)(\omega-Q_+)+\Gamma^2 } \, ;&
                  -V < \omega < 0
                  \end{array} \right.
           \; . \;
\end{eqnarray}
Combining Eqs.~(\ref{Zvariation}) and (\ref{Ddef}),
one arrives at the following expression for the
$Z$-functional:
\begin{eqnarray}   \label{afterint}
 \ln Z/Z_0 = - \int dt \int_0^{Q_-} dQ_-  \int \frac{d \omega}{2
 \pi} D^{--}(\omega) e^{i \omega 0^+} \nonumber \\
  + F(Q_+) \, ,
\end{eqnarray}
where $F(Q_+)$ is a yet unknown function depending only on $Q_+$
and the explicit time-dependence of $Q_\mp$ has now been restored.
This procedure results in the loss of non-local in time, dissipative
terms, see Ref.\cite{yuanderson}, which are accessible via perturbative
expansions or a generalised Wiener-Hopf method. Such terms are
only important for tunnelling between different molecular configurations
and will be addressed in the long version \cite{inpreparation}.
In the above $Z_0$ is responsible for the dynamics of the bosonic mode
without the electron-phonon coupling and is
given by the functional integral
\begin{eqnarray}   \label{ZQ}
 Z_0 = \int {\cal D} Q_- {\cal D} Q_+ e^{i S_0[Q_\mp]} \, ,
\end{eqnarray}
with the free action
\begin{eqnarray}
 S_0[Q_\mp] = \frac{1}{2 \lambda^2} \int dt \Big[
 M (\dot{Q}_-^2 - \dot{Q}_+^2) -
 K(Q_-^2-Q_+^2) \Big]  \, .
\end{eqnarray}
The energy integration in Eq.~(\ref{afterint}) results in two
contributions:
\begin{eqnarray}
 \ln Z/Z_0 = - i \int dt \int_0^{Q-} d Q [ n_d(Q) &+& i f(Q,Q_+)]
\nonumber \\ &+&  F(Q_+) \, .
\end{eqnarray}
The first contribution is purely imaginary and is related to the occupation
probability of the dot level,
\begin{eqnarray}              \label{population}
 n_d(Q) = \frac{1}{2 \pi}
 \Big[\cot^{-1}(Q/\Gamma)
 + \cot^{-1}((Q-V)/\Gamma) \Big] \, .
\end{eqnarray}
The second contribution is real and is given by
\begin{eqnarray}        \label{ffunc}
 f(Q_-,Q_+) = \int_{-V}^0 \frac{d \omega}{2 \pi} \Big[
 \frac{\omega-Q_+}{(\omega-Q_-)(\omega-Q_+) + \Gamma^2}
 \nonumber \\
 - \frac{\omega-Q_-}{(\omega-Q_-)^2+\Gamma^2} \Big] \, .
\end{eqnarray}
Performing analogous computation the field $Q_+$
(or on symmetry grounds), one identifies the missing
function as
\begin{eqnarray}
 F(Q_+) = - \int^{Q_+}_0 d Q [n_d(Q) + i f(Q,0)] \, .
\end{eqnarray}
Hence the effective action:
\begin{eqnarray}         \label{effaction}
 S &=& S_0 - i \int dt
\Big\{ \int_0^{Q_-} d Q [ n_d(Q) + i f(Q,Q_+)]
 \nonumber \\
&-& \int_0^{Q_+} d Q [n_d(Q) + i f(Q,0)] \Big\} \, .
\end{eqnarray}
The next step is to minimise the action, which yields the system
of coupled equations ($\eta=\pm$):
\begin{eqnarray}               \label{eqsyst}
K Q_\eta/\lambda^2 + n_d(Q_\eta) - i \eta f(Q_\eta,Q_{-\eta}) =0
\, .
\end{eqnarray}
Let us discuss a special class of solutions which satisfy
$Q=Q_-=Q_+$. Function (\ref{ffunc}) vanishes in this case, so all
these solutions are real and satisfy the following equation (we
restored the energy offset $\Delta$ of the dot level)
\begin{equation}      \label{findQe}
 -K Q/\lambda^2 = n_d(Q+\Delta) \, .
\end{equation}
We set $V=0$ first. It is easy to see that for
$\Delta<0$ and $\Delta \gg 1/K$ there is only one solution of this
equation, see Fig.~\ref{potplot}.
The former situation corresponds to the dot level being
placed below the Fermi edges of the leads.
In this case the equilibrium position of the
phonon mode lies below the Fermi energy. For large and positive
gate voltage $\Delta \gg \lambda^2/K$ the dot level is effectively
attracted by the Fermi sea and the dot energy level is slightly
lowered. Unless the slope of the rhs of Eq.~(\ref{findQe}) at
$Q=0$ is smaller than $-K/\lambda^2$, which corresponds to $\pi
K/\lambda^2 > \Gamma^{-1}$, there is a third regime when the line
$-K Q/\lambda^2$ crosses $n_d(Q+\Delta)$ three times. In this case
the effective potential for the boson mode acquires two dips, see
Fig.~\ref{potplot}, and the system becomes bistable. A similar
bistability in a different (negative-$U$) dot model has recently
been reported in Ref.\cite{bratkovsky}.
The non-equilibrium case is even richer.
The population probability
$n_d$ now  possesses two steps as a function of energy. This opens
the way to tristable regimes as Eq.~(\ref{findQe}) can have five
solutions, see Fig.~\ref{potplotV}. As in the equilibrium
situation multiple solutions are only possible for sufficiently strong
electron-phonon coupling, $\pi K/\lambda^2 < \Gamma^{-1}$.
Remarkably, the bistability phenomenon should be visible in the
non-equilibrium transport even without any adjustments of the gate
voltage. Starting from the equilibrium system with $\Delta=0$ and
increasing the bias voltage, see Fig.~\ref{IV}, one inevitably
crosses the bistability regime, which reveals itself as a
hysteresis in the $I-V$ diagram.

For $Q_-=Q_+$, the correction to the effective action,
(\ref{effaction}), can be interpreted as an effective
non-equilibrium free energy of the phonon mode. We looked for but
could not find time-inversion breaking minima with $Q_- \neq Q_+$.
Thus we can confirm that for the system under consideration the
`quantum Onsager reciprocity principle', put forward in
Ref.~\cite{coleman}, is correct. We stress that the Keldysh action
approach does not require any external conjectures to hold and
allows, at least in principle, for solutions with $Q_- \neq Q_+$,
which would be at odds with Ref.~\cite{coleman}. Also, even around
the time-invertion symmetric minima, the (slow) phonon dynamics
involves fluctuations of two independent but coupled fields
$Q_\mp$. (In equilibrium all these complications disappear as
$Q_\mp$ are then simply decoupled.)

With applications to (single-wall) carbon nanotubes in mind,
we now briefly discuss the case of interacting leads (in equilibrium).
The adequate description in low-energy
sector is then given by the Luttinger liquid (LL) model \cite{haldane,book}.
To describe the leads we employ the open boundary
bosonization procedure.
In this approach the electrons can be
thought of as chiral particles living on the whole real axis. The
tunnelling onto and from the dot level takes place at $x=0$ in
both leads. The Hamiltonian has the simplest form
in terms of the Bose fields $\phi_{1,2}$ which describe the
collective low-density plasmon excitations in the corresponding
leads,
\begin{eqnarray}      \nonumber
 H_0=H_0[\psi_1] + H_0[\psi_2] = \frac{1}{4 \pi} \int dx \left[
 (\partial_x \phi_1)^2 + (\partial_x \phi_2)^2 \right] \, .
\end{eqnarray}
The local (at $x=0$) field operators in this representation are
then given by \cite{book},
\begin{eqnarray} \label{psidef}
 \psi_i(t) = (2 \pi a_0)^{-1/2} e^{i \phi_i(t) / \sqrt{g}} \, .
\end{eqnarray}

This equation contains the Luttinger liquid parameter $g$, which
is related to the interaction strength $U_0$ via  $g=(1+ 4
U_0/\pi)^{-1/2}$ \cite{haldane}, and the lattice constant of the
underlying lattice model $a_0$.
As a next step we substitute (\ref{psidef}) into (\ref{H0}) and
re-write the dot level operators in terms of spin-1/2 operators,
$S_y = -i(d^\dag - d)/2$, $S_z = d^\dag d - 1/2$. After applying a
canonical transformation, $U=\exp[ i S_z (\phi_1(0) + \phi_2(0))/2
\sqrt{g}]$, to the Hamiltonian of the system, $H' = U^\dag H U$,
and introducing new fields $\phi_\pm = (\phi_1 \pm
\phi_2)/\sqrt{2}$ we arrive at the following Hamiltonian:
\begin{eqnarray}            \label{kondoham}
 H' = H_0[\phi_\pm] &+& Q S_z -i  \gamma \sqrt{2/\pi a_0} S_y
 \cos[\phi_-(0)/\sqrt{2 g}] \nonumber \\
  &-& \nu \sqrt{2/g} S_z \partial_x \phi_+(0) \, ,
\end{eqnarray}
where we introduced an additional coupling constant $\nu$.
Transformed into this form the Hamiltonian is closely related to the Kondo
problem \footnote{A related system of only one lead coupled to a
purely \emph{static} localised level was recently studied in Ref.
\cite{furusakimatveev}.}. The field $Q$, which
is assumed to be adiabatically slow in
comparison to electronic degrees of freedom, plays the role of the
magnetic field applied to the localised spin $\vec{S}$. The
population probability $n_d$ is now related to the magnetisation
$n_d = \langle S_z \rangle + 1/2$. Correspondingly, the magnetic
susceptibility gives us the derivative of the $n_d$ with respect
to the boson coordinate $Q$.

The special point $\nu=0$ is analogous to the Toulouse limit in
the Kondo problem. Moreover, at $g=1/3$ the Hamiltonian coincides
with that of the four-channel Kondo problem, see Ref.~\cite{book}.
In order to analyse the magnetic susceptibility we proceed along
the lines of Ref.~\cite{book}. The large time asymptotics of the
spin-spin correlation function turns out to be
 $\langle T S_z(t) S_z(0) \rangle = C t^{-2g}$,
$C$ being a numerical non-universal constant. Therefore the
temperature (and hence energy) dependence of the magnetic
susceptibility $\chi = dn_d/d Q$ is
(after the appropriate continuation to imaginary time):
\begin{eqnarray}
 \chi = C^2 \int_0^{1/T} d \tau \langle S_z(\tau) S_z(0) \rangle \sim
 C_1 \, T^{2g-1} \, ,
\end{eqnarray}
where $T$ is the
temperature and $C_1$ is another non-universal constant.
The last formula has very important consequences for the existence
of multistable regimes. In the case of strongly interacting
electrodes, when $g<1/2$, $\chi$ is divergent and the slope of
$n_d$ is infinite.
Hence the multistable regimes are always possible when the gate
voltage is adjusted in appropriate way and there is no critical
electron-phonon coupling.
However, $\chi$ vanishes
as the temperature or the relevant energy scale approaches zero
for weak interactions in the leads, for $g>1/2$.
So, for all $1/2<g<1$ there is a threshold in $\lambda$
above which the multistable regimes emerge.

To conclude, we have investigated transport properties of a
quantum dot containing one level coupled to a local phonon mode.
In the case of non-interacting leads we derived the effective
action for the phonon, the analysis of which reveals the existence
of a bistable regime in the equilibrium and a tristable regime for
finite bias voltage.
For strongly interacting leads, the
conditions on the electron-phonon coupling needed to enter
multistable regimes are relaxed.
One obvious avenue for further
research is the phonon dynamics close to the bistability where the
dissipation rate changes from Ohmic to super-Ohmic at  finite bias
voltage \cite{inpreparation}.
The bulk interaction effects merit further investigation
(so, the $g=1/2$ model is exactly solvable by re-fermionization,
even at finite $V$.)
The most interesting emerging
theoretical question is about the viability of time-inversion
breaking solutions in similar systems, this question is currently
open.

This work was partly supported by the EPSRC of
the UK under grants GR/N19359 and GR/R70309 and the
EC training network DIENOW.

\bibliography{prb3}
\end{document}